\documentclass[
 paper=A4,pagesize=automedia,fontsize=10pt,
 DIV=22, BCOR=0mm,
 parskip=never,
 titlepage=false,
 abstract=false,
 ]{scrartcl}
\usepackage{lmodern}
\usepackage{graphicx}
\usepackage[utf8]{inputenc}
\usepackage[doublespacing]{setspace}
\usepackage{placeins}
\usepackage{amsmath,amssymb}
\usepackage{nicefrac}
\usepackage[FIGTOPCAP]{subfigure} \subfiglabelskip=0pt
\usepackage[labelfont=bf, font={small, stretch=1.5}]{caption}
\DeclareCaptionLabelSeparator{bar}{ $\mid$ }
\usepackage[superscript]{cite}
\usepackage[colorlinks = true
            ,allcolors = blue
            ]{hyperref}
\newcommand{\sel}{$s_{\text{EL}}(\lambda)$}
\newcommand{\spl}{$s_{\text{PL}}(\lambda)$}
\newcommand{\diff}{\mathrm{d}}
\newcommand{\reffig}[2][]{\hyperref[#2]{Fig.~\ref{#2}\textcolor{blue}{#1}}}
\newcommand{\refeq}[1]{\hyperref[#1]{(\ref{#1})}}
\begin{document}
\title{Real-time beam-shaping without additional optical elements}
\author{Felix Fries\thanks{felix.fries@iapp.de},
	Markus Fr\"obel,
	Pen Yiao Ang,
	Simone Lenk,
	Sebastian Reineke\thanks{sebastian.reineke@iapp.de}\\
	\and \small Dresden Integrated Center for Applied Physics and Photonic Materials (IAPP),\\ \small Technische Universit\"at Dresden, N\"othnitzer Strasse 61, 01187 Dresden, Germany}
\date{}%
\maketitle
%
%\linenumbers
% Abstract:
\textbf{
Providing artificial light and enhancing the quality of the respective light sources is of continued interest in the fields of solid state-, condensed matter, and semiconductor physics. A lot of research has been done to increase luminous efficiency, lifetime and colour stability of such devices. However, the emission characteristics of a given light source don't necessarily comply with today’s, often sophisticated applications. Here, beam-shaping deals with the transformation of a given light distribution into a customized one. This is achieved by secondary optical elements often sporting elaborate designs, where the actual light source takes up only a small fraction of the system's volume. Such designs limit the final light source to one permanent operation mode, which can only be overcome by employing mechanically-adjustable optical elements. Here, we show that organic light-emitting diodes (OLEDs) make real-time regulation of a beam-shape possible, without relying on secondary optical elements and without using any mechanical adjustment. For a red light-emitting two-unit OLED architecture, we demonstrate the ability of continuous tuning between strongly forward and strongly sideward emission, where the device efficiency is maintained at an application relevant level ranging between 6-8\% external quantum efficiency (EQE) for any setting chosen. In combination with additional optical elements, customizable and tunable systems are possible, whereby the tuning stems from the light source itself rather than from the secondary optics.
	}%

% Maint text:
\begin{figure}[!th]
	\centering
	\includegraphics[width=\textwidth]{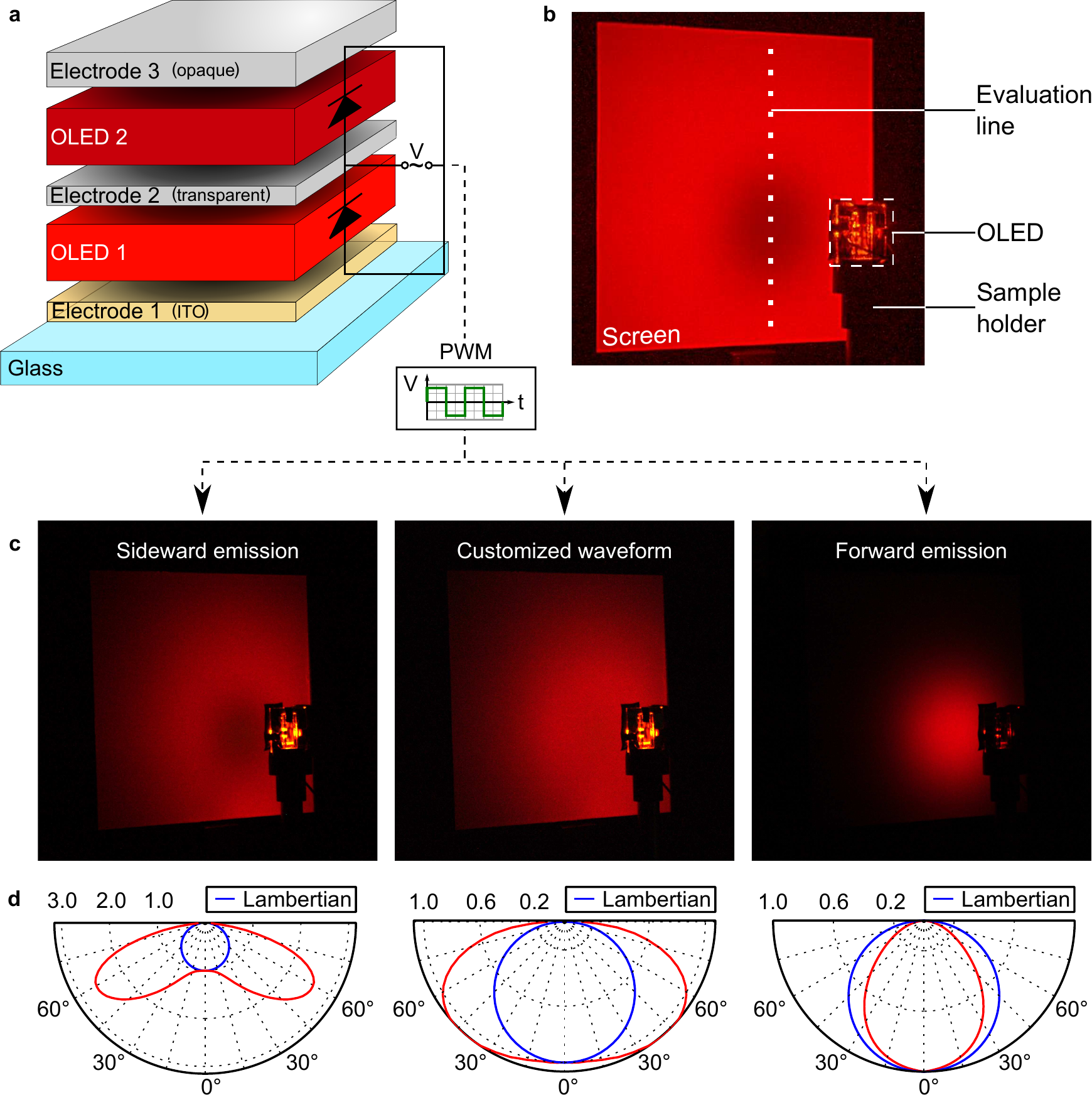}
	\caption{\textbf{Working scheme of active beam-shaping OLEDs.} a) Two stacked OLED units, electrically connected for AC operation. b) Experimental setup for analysis of the brightness distribution on a planar, two-dimensional surface. The OLED and the screen have to be parallel to each other. The evaluation of the brightness is done in the vertical direction as indicated by the dotted line. c) Using PWM, different AC-signals can be applied to the device resulting in either a ring shaped beam (left), broad homogeneous illumination (centre) or a focused spot (right). d) Respective irradiance as a function of the viewing angle for the conditions shown in c), measured using a spectro-goniometer.}
	\label{fig:Setup_etc}
\end{figure}

Beam-shaping is a prominent topic in various fields of optics. In laser physics, optical elements are commonly used to transform the Gaussian-shaped TEM$_{00}$ mode into various spatial distributions that are defined by the application's need \cite{Dickey2014}. In such developments, sophisticated theoretical analysis and elaborated designs of phase plates are utilized \cite{Wei2014}. Inorganic LEDs are point-sources in most settings and are therefore easily and effectively combined with secondary beam-shaping systems \cite{Fournier2011}, metallic nanostructures or optical nanoantennas \cite{Lozano2016}. OLEDs, being area light sources with broad-band emission \cite{Reineke2013a} and often Lambertian emission characteristics, have not been subject to serious beam-shaping designs. Rather, efforts in developing elaborate optical concepts aim to increase the outcoupling efficiency, which for planar devices only ranges between 20-30\% \cite{Reineke2013a, Gather2015}.
A few studies focus on OLED-based beam-shaping by moving away from the planar architecture \cite{Lee2014}, employing diffractive gratings to achieve a very directional emission \cite{Zhang2014}, or by reducing the size of OLED pixels such that, when combined with microlens arrays, their output resembles that of a point source.\cite{Melpignano2006} What all of the above concepts have in common is the fact that only one final light distribution is realized, which is defined by the optical interplay between light source and secondary optical elements. To circumvent this limitation, mechanically tuneable elements are necessary and result in a setup that is much more complicated \cite{Dickey2014}. Such complicated setups are undesirable particularly when it comes to applications like car lights, home illumination and information displays. Here, simple and cost-effective
solutions with a small system footprint are preferred.

We have revisited the possibilities of OLED technology and, through careful optical design, developed an OLED-based area light source with high efficiency, good colour stability and most importantly, significant angular tuneability of the emitted light. Neither passive optical elements nor movable parts are needed and the beam shape can be altered continuously during operation. This is why we refer to our concept as \textit{active} beam-shaping. 

\subsection*{Efficient OLEDs with variable angular emission pattern}

\reffig[a]{fig:Setup_etc} shows the general device layout employed in this study (for details of the OLED architecture, refer to Fig. S1 in the Supplementary Information). Two sub-units are stacked vertically in a bottom-emitting configuration, meaning that the device emits through the glass substrate. Compared to conventional OLEDs, this device shows an increased overall thickness of about 800 nanometres, which is needed to realize the desired optical effects. Here, electrically-doped charge transport layers allow for a free design of the optical system without sacrificing electrical performance \cite{Furno2012}. These units are connected in a way that either OLED 1 or OLED 2 is addressed. This driving concept, referred to as AC/DC driving, has been used before for colour tuning \cite{Froebel2015} utilizing an emission ratio control of the units through pulse-width modulation (PWM). For visualization of the tuneable emission, a setup is chosen, where the OLED’s emission is projected on a planar white screen, as illustrated in \reffig[b]{fig:Setup_etc}. Three different driving scenarios are illustrated in \reffig[c]{fig:Setup_etc}: Addressing only OLED 1, a central spot is observed on the projection screen. On the contrary, the emission pattern changes to a distinct ring shape in the case of addressing only OLED 2. Finally, by combining time-averaged emission from OLED 1 and OLED 2 via PWM, a customized emission pattern can be realized. The switching ratio in such device is only limited by the time needed to reach a steady state in the actively emitting unit, which is in the 5~$\mu s$ range for phosphorescent OLEDs \cite{Reineke2008} and for fluorescent ones even some nanoseconds.\cite{Kasemann2011}. High speed beam-shaping solutions in literature prove switching velocities in the same regime or below (kHz - MHz).\cite{Bechtold2013, Schenk2014, Cheng2015}
Characterization of the units' emission profile is done in a spectro-goniometer, where a detector measures the angular-dependent emission intensity. % by moving with constant distance from 0$^\circ$ to 90$^\circ$ around the device, which is sitting in the centre of the configuration.
The results are shown in \reffig[d]{fig:Setup_etc}. The sideward-emitting OLED 2 has its emission intensity maximum at 56$^\circ$, showing a 2.7-fold increase over its emission at 0$^\circ$. At this angle, the forward-emitting unit provides only 30\% of its own 0$^\circ$ intensity, indicating a very clear separation of the two extreme operation modes. A more detailed stepping of the PWM-controlled angular emission profiles is summarized in the Supplementary Information, Fig. S2 and S3. Additionally, an animated Graphics Interchange Format (GIF) summarizing projections on the screen for a sweep of different driving conditions is available in the Supplementary Material. 

\reffig{fig:Device_characteristics} contains the key performance characteristics of the red, active beam-shaping OLEDs. \reffig[a]{fig:Device_characteristics} shows current density (j)-voltage (V)-forward luminance characteristics of both units independently. Very steep jV-characteristics are observed despite the very large overall device thickness (about 800 nm), which is made possible by comprising highly-conductive transport layers \cite{Huang2002}. The forward luminance for OLED 2 is, of course, lower as it is designed to emit efficiently to high angles of observation. EQE versus j is shown for both units independently in \reffig[b]{fig:Device_characteristics}, having maximum values of 6.6\% for forward emitting OLED 1 and 8.2\% for sideward emitting OLED 2, respectively. The EQE of the forward emitting unit is limited here, because the photoluminescence quantum yield of the emitter used (Ir(piq)$_3$) is only approximately 50\% \cite{Reineke2007, Suzuki2009}. \reffig[c]{fig:Device_characteristics} shows the electroluminescence (EL) spectra of OLED 1 and OLED 2 for their respective maximum intensity emission, which are spaced by 54~nm.

\begin{figure}[!t]
	\centering
	\includegraphics[width=\textwidth]{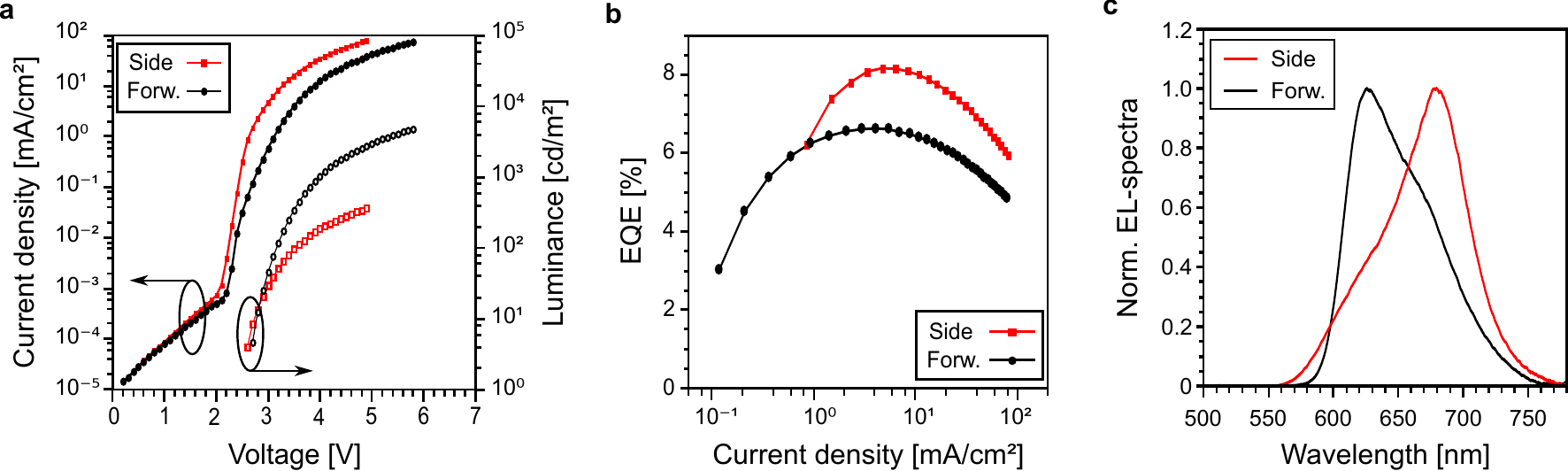}
	\caption{\textbf{Electro optical device characteristics.} a) The current density as a function of the driving voltage is shown in black, the respective forward luminance in red. b) The resulting EQE for both units. c) EL-spectra for both the side and the forward emission unit, at their emission maximum. The peaks show a spectral distance of 54~nm.}
	\label{fig:Device_characteristics}
\end{figure}

\subsection*{Optical design of the beam-shaping OLEDs}
The final OLED stack for effective beam-shaping should show very anisotropic emission patterns for the two units employed. Therefore, one needs to analyse the explicit expression of the spectral radiant intensity (SRI) per unit area $I(\lambda,\theta)$, as given by Furno \textit{et al.} \cite{Furno2012}:

\begin{equation}\label{eq:SRI}
	I(\lambda, \theta) = \frac{hc}{\lambda} \cdot \frac{I}{e} \cdot \gamma \cdot s_{\text{EL}}(\lambda) \cdot \underbrace{\eta^\star_\text{rad}(\lambda)\cdot \frac{P_\text{out}(\lambda, \theta)}{F(\lambda)}}_\text{Cavity-Mode},
\end{equation}

with the wavelength $\lambda$, the viewing or emission angle $\theta$, Planck's constant $h$, the speed of light in vacuum $c$, the current $I$, the elementary charge $e$, the charge carrier balance factor $\gamma$, and the EL spectrum of the embedded emitter \sel, which normally gets identified with the directly-measurable photoluminescence (PL) spectrum \spl \cite{Chen2007}. The second-to-last factor $\eta^\star_\text{rad}(\lambda)$ is the effective radiative efficiency of the emitter, which is  the ratio of the radiative and the total decay rate in the presence of a microcavity \cite{Furno2012, Chen2007, Neyts1998}. The outcoupled power spectrum per unit solid angle $P_\text{out}(\lambda, \theta)$ is the only factor that contains an explicit angular dependency and is divided by the Purcell-Factor $F(\lambda)$ \cite{Purcell1946}. Thus, the factor $\frac{P_\text{out}(\lambda, \theta)}{F(\lambda)}$ gives the emission affinity for a photon of a given wavelength $\lambda$ under an angle $\theta$ within the microcavity. Accordingly, for a given emitter \spl, only the last two factors in equation~\refeq{eq:SRI} define the final out-coupled spectra of the electroluminescent device. From an engineering point of view, it is the total thickness of the cavity, the reflectivity of the mirrors and the positioning of the emitting layers that define these two factors. Hence, they are referred to as cavity mode. Influences of the microcavity on the final out-coupled spectrum of OLEDs have been widely analysed before and even preferential side emission has been observed previously \cite{Neyts1998, Purcell1946, Furno2012, Mladenovski2009}. However, none of these studies have investigated a possible sideward emission with respect to maximum efficiency.

Key influences on the cavity mode need to be considered to allow for effective beam-shaping as demonstrated here. First, the thickness of the transparent contacts in the device have a determining role. In analogy to a Fabry-P\'erot interferometer, a higher reflectance of the mirrors leads to a higher finesse, resulting in a narrower cavity mode and, consequently, a reduced full width at half maximum (FWHM) of the emission spectra \cite{Becker1997}. Those arguments hold true for both the ultra-thin wetting layer metal electrodes and metal oxides like indium tin oxide, which are used in the device layout (cf. \reffig[a]{fig:Setup_etc}) \cite{Schubert2013, Lenk2015}.

Moreover, a change of the cavity length, meaning the resonance wavelength \cite{Svelto2010}, strongly influences the emission of OLEDs \cite{Chen2007, Becker1997, Bruetting2013}. In this study, the thickness of the device is adjusted through thickness changes of the highly conductive and transparent doped transport layers. Importantly, OLEDs are different than passive cavities, because the active layer where EL originates, the emission layer (EML), accounts only for a small fraction of the total cavity thickness (around 20 nm typically) and can be placed freely within the cavity. Thus, changes can be achieved for a given cavity length and variable EML position. For an unambiguous description, it is necessary to consider the distribution of the electromagnetic field in layered media \cite{Furno2012, Fries2016, Sipe1987}. OLEDs enclosing one field maximum are denoted as OLEDs of first order. Analogue OLEDs of second order show two maxima. The latter architecture allows various ways of positioning of the EML: in one of the two field maxima (optical maximum) or even in the minimum (optical minimum). 

\begin{figure}[!t]
	\centering
	\includegraphics[width=\textwidth]{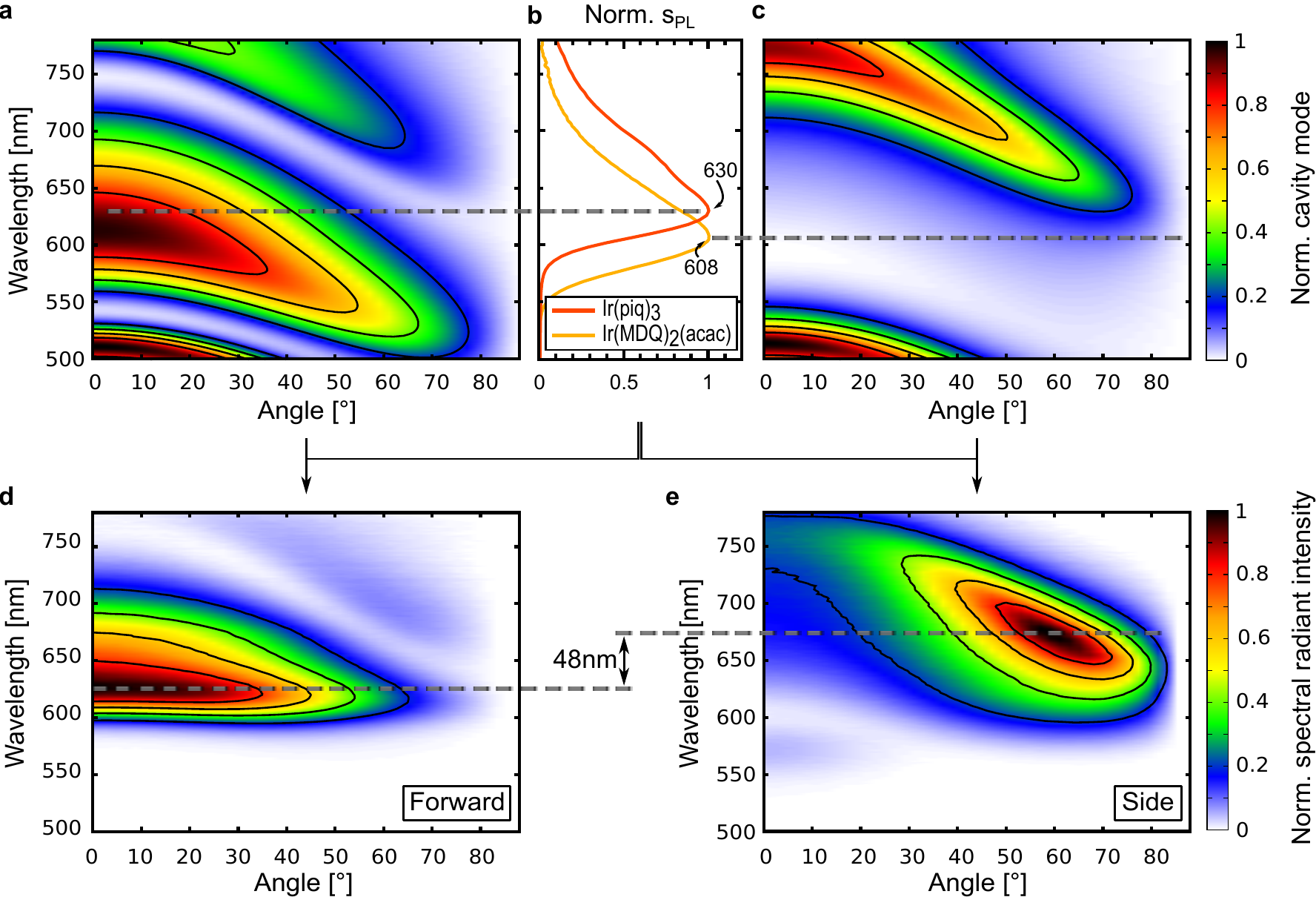}
	\caption{\textbf{Simulation of the cavity mode and the SRI for beam-shaping OLEDs.} Calculated cavity modes are shown for an OLED in the optical maximum (a) and the optical minimum (c), respectively. b) shows the normalized PL spectra of the red emitters used. The peak intensities of both emitters are marked in the cavity modes to clarify the choice of the emitter selection. Multiplying the cavity modes with the PL-spectra explains the measured SRI as shown in d) for forward emission and e) for sideward emission. The spectral difference between the maxima is 48~nm.}
	\label{fig:Cavity_mode_and_spectra}
\end{figure}

Simulations demonstrate that OLEDs of higher order show narrower cavity modes, which bend to shorter wavelengths with increasing angles (compare \reffig[a]{fig:Cavity_mode_and_spectra} and \reffig[c]{fig:Cavity_mode_and_spectra}). The curvature follows the cosine-like behaviour, typical for a Fabry-P\'erot interferometer \cite{Svelto2010}. This effect is used for the desired angularly-confined emission. Primarily forward emission emerges from placing the EML close to the optical maximum. In our study, the cavity length is optimized for the dark red emitter Ir(piq)$_3$. The resulting experimentally measured spectra (cf. \reffig[d]{fig:Cavity_mode_and_spectra}) show maximum intensity at 0$^\circ$ and 626~nm, dropping to half intensity at 48$^\circ$. Outcoupling of another branch of the cavity mode gets visible at higher angles, but can be ignored due to its low intensity.

In order to realize a condition where the 0$^\circ$ emission is suppressed effectively, one has to place the EML in the respective optical minimum of the cavity (cf. Fig. S1, Supplementary Information). The calculation of the respective cavity mode (cf. \reffig[c]{fig:Cavity_mode_and_spectra}) shows that indeed one can expect emission at higher angles, around 50-80$^\circ$. A comparison of \reffig[a]{fig:Cavity_mode_and_spectra} and \reffig[c]{fig:Cavity_mode_and_spectra} indicates an offset between the respective modes, with the sideward emission shifted to a longer wavelength. To compensate for this effect, another emitter is chosen, namely the red emitter Ir(MDQ)$_2$(acac). Its PL-spectrum resembles the one of Ir(piq)$_3$, but is blue-shifted about 22~nm (see \reffig[b]{fig:Cavity_mode_and_spectra}). This approach may seem counter-intuitive but it is one of the most important design steps towards colour stability between forward and sideward emission. The experimentally obtained SRIs are shown in \reffig[d) \textcolor{black}{and} e)]{fig:Cavity_mode_and_spectra}. Because of the aforementioned mode curvature, the resulting SRI for sideward emission peaks at higher viewing angles of around 60$^\circ$. These SRIs of forward and sideward emission are compared to simulated values, calculated by multiplying the cavity modes with the respective PL spectra of the emitters, in Fig. S4 (Supplementary Information). The difference in wavelength between the experimental maxima in forward and sideward emission is 48~nm. This is slightly smaller than the value obtained from the spectra in \reffig[c]{fig:Device_characteristics}, which is due to the fact that the latter refers to the maximum of the integrated radiant intensity.
Without compensation through emitter choice, the peak differences would be as high as 66~nm, as simulated for the case of the emitter Ir(MDQ)$_2$(acac). With the right emitter molecules at hand, this colour difference can be further minimized.

\subsection*{Various planes of observation}

Both the simulation and the measurement using a spectro-goniometer follow a radial symmetry with the OLED in the centre of rotation, mimicking a spherical surface of observation. However, for many conceivable applications this does not represent a very realistic model. Therefore, as introduced earlier, the analysis of the beam-shaping performance is carried out using a planar, diffusive screen (see \reffig[b]{fig:Setup_etc}). A subsequent image analysis allows us to obtain the brightness distribution. The results of spectro-goniometer and imaging screen measurements are in direct correlation to each other via an azimuthal transformation, as sketched in \reffig[a]{fig:Geometry}. This case, where the projection source is in the centre of the sphere, is called gnomonic projection. Assuming the OLED to be a point source and the aperture of the spectrometer as d$A$ at the distance $R$, the corresponding area on the screen is d$A^\prime$, which is at a distance $r$ from the spheres north pole ($\theta_\mathsf{north~pole} = 0^\circ$). Because the diameter of d$A$ is much smaller than $R$, it can be interpreted as the surface element in spherical coordinates and analogue d$A^\prime$ in planar coordinates. Simple calculation yields for the irradiance~$E$:

\begin{equation}\label{eq:Irradiance-Trafo}
        E_\text{Goniometer} = \frac{1}{\cos^3 \theta} \cdot E_\text{Screen}.
\end{equation}

If the screen is at a distance $R^\prime\neq R$ from the OLED, equation~\refeq{eq:Irradiance-Trafo} contains an additional factor $\left(\nicefrac{R^\prime}{R}\right)^2$. However, by normalizing the spectra, this factor cancels again. It is worth mentioning that the surface element in spherical coordinates is not of constant size, but dependent on the azimuthal angle $\theta$. Yet, the spectrometer comprises a constant aperture. This situation can be modelled by cylindrical coordinates. In this case the transformation factor becomes $\nicefrac{1}{\cos^2 \theta}$. However, projecting the respective surface element on a two dimensional plane does not give a filled surface. Hence, the true transformation should lie between those two models. This transformation points out the importance that the integrated radiant intensity for sideward emission has to be strongly enhanced (here, a factor 2.7 to normal emission,  cf. \reffig[d]{fig:Setup_etc}). Otherwise a hollow image, as shown in \reffig[c]{fig:Setup_etc} for the sideward emission, cannot be realized.

\begin{figure}[!t]
	\centering
	\includegraphics[width=\textwidth]{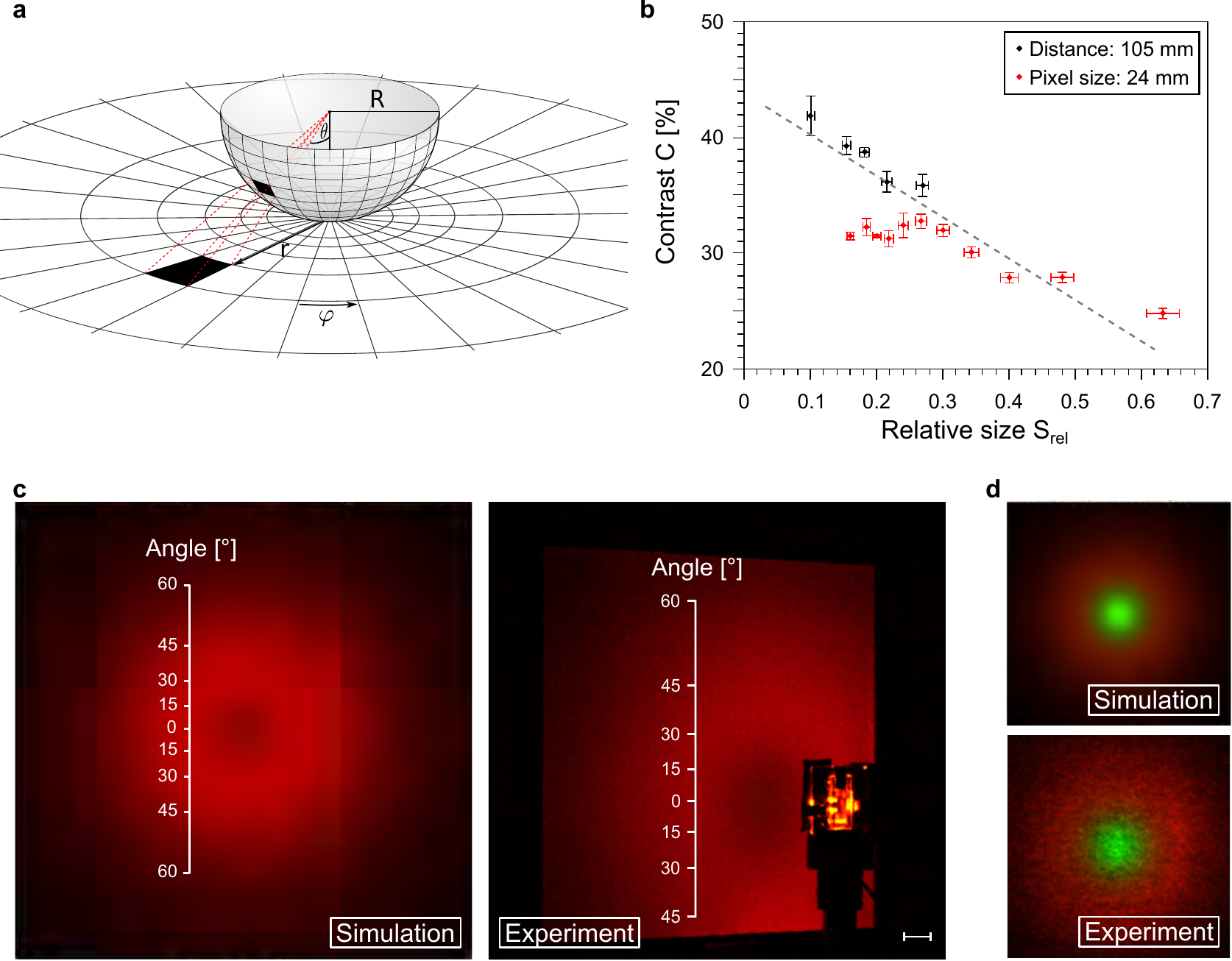}
	\caption{\textbf{Observers position and contrast.} a) Scheme of the transformation needed to compare the measurements obtained on a spectro-goniometer and on a planar screen. b) The contrast is shown with respect to the relative size $S_{\mathrm{rel}}$, which is the ratio of the size of the OLED to its distance from the screen. Black points were obtained by keeping the distance constant; red points by keeping the size constant. The smaller $S_{\mbox{rel}}$, the better the contrast. The error bars represent the statistical confidence interval of 1$\sigma$. c) With the methods explained herein, it is possible to predict the beam-shape from the spectro-goniometer data. Here, the results for the red OLEDs developed in this contribution are shown. Scale bar is 10 mm. d) To confirm that not only the brightness distribution but also the colours are calculated correctly, samples were made, that emit greenish in the centre of the spot and red to the edges.}
	\label{fig:Geometry}
\end{figure}

Using the above transformation, the beam shape can be predicted if the spectral information is provided either by simulation or the goniometer measurement, which is of great importance for further optimization. Using CIE2006 colour matching functions and a subsequent linear transformation to RGB-colour space, a displayable picture of the beam's shape is achieved \cite{CIE2006, Hunt2004}. In \reffig[c]{fig:Geometry} a calculated image is compared to a photograph. Both the brightness distribution and the colour are well reproduced.
To prove the reliable reproduction of colours, OLEDs were build that emit green in the centre of the spot and red to the sides (cf. \reffig[d]{fig:Geometry}). The brightness-colour distribution is still well reproduced. Further information on this transformation is given in the Supplementary Information (cf. Fig.~S5 and S6).

\begin{figure}[!t]
	\centering
	\includegraphics[width=\textwidth]{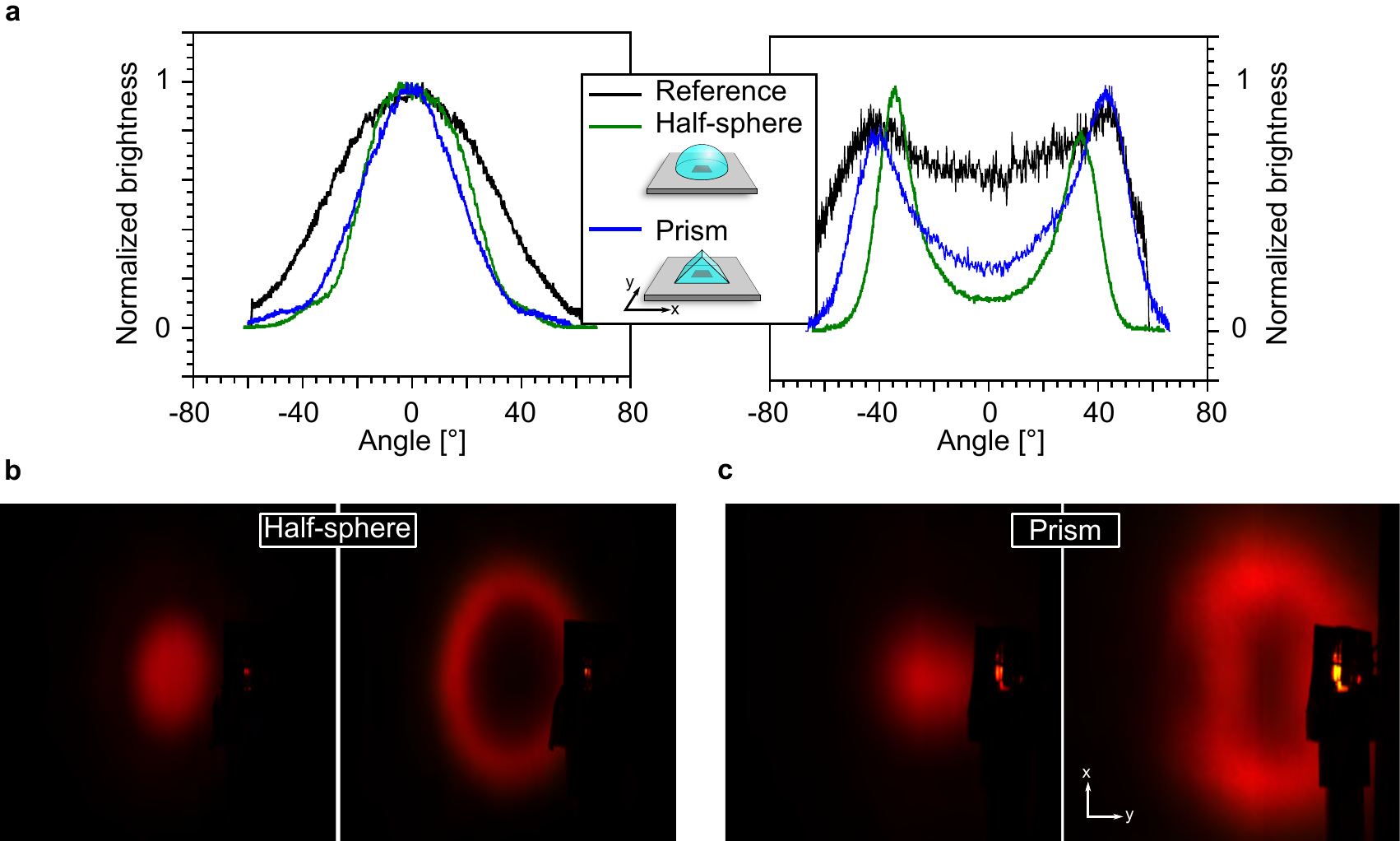}
	\caption{\textbf{Adding secondary optical elements to active beam-shaping OLEDs.} a) The normalized brightness distribution along the evaluation line for forward (left) and sideward (right) emission are shown with and without additional optical elements placed on top of the OLED. The images on the projection screen obtained with a half sphere b) and prism c), respectively.}
	\label{fig:Secondary_optics}
\end{figure}

There remains an influence on the shape of the beam that has not yet been discussed. In the considerations above, the OLED has been treated as a point source. This gives rise to question the limits in which this assumption is right. There are two general ways to investigate this behaviour: either the size of the OLED or its distance to the screen must be varied. In the following discussion, the ratio of the size of the pixel $l_\text{dia}$ and the distance $d_\text{scr}$ will be analysed, wherein the size denotes the length of the diagonal of the square pixel. This ratio will be referred to as the relative size $S_{\text{rel}}=\nicefrac{l_\text{dia}}{d_\text{scr}}$. For large values of $S_\text{rel}$, rays emitted under various angles at different points of the source will hit the screen on the same spot, blurring the contrast. Experiments are carried out with beam-shaping OLEDs of $l_\text{dia}=28$~mm pixel size. Using shadow-masks, the pixel size could be varied stepwise at a constant distance to the screen of 105~mm. As expected, the resulting contrast increases with decreasing relative size, as shown in black in \reffig[b]{fig:Geometry}. More surprising is the result for a constant pixel size of $l_\text{dia}=24$~mm. For relative sizes larger than 0.3 the values add nicely to the ones in black, but around 0.3 the curve is kinked and starts to drop slightly again towards smaller $S_{\text{rel}}$ values. Because of the conservation of \'etendue in free space \cite{Winston2005}, this cannot be a geometrical effect. Rather, it is based on greater noise in the spectra with increasing distance, due to longer exposure times of the camera. Assuming $I$ the maximum intensity, $i$ the intensity at the centre of the spot, $\Delta I$ the contribution of a constant noise, $C$ the contrast without noise, and $\tilde{C}$ the contrast with noise, one gets:

\begin{equation}
\tilde{C} = \frac{(I+\Delta I)-(i+\Delta I)}{I+\Delta I}=\frac{I-i}{I+\Delta I}\leq \frac{I-i}{I}=C.
\end{equation}

Clearly, a non-zero constant noise of our camera system will limit the contrast values, as soon as the measured intensities become comparable to the noise floor. Increasing the driving current would counteract to this effect, but to make the interpretation of the data as clean as possible, it was kept constant over the whole measurement. These results also show that a definite prediction of the beam shape on the screen is only possible within the assumption of ideal measurement conditions.

\subsection*{Combining active and passive beam-shaping}

So far we have presented the possibilities offered by active beam-shaping using the planar, bare OLEDs. At the same time, this system offers to be used as a novel, adjustable platform in combination with additional, secondary optics. As an example, we will discuss possible effects using primitive optical elements, in this case a half-sphere and a prism, to demonstrate the variability of this promising concept.

The difference in the refractive indices especially between the substrate glass and air leads to total internal reflections. Placing a glass half-sphere on top of the substrate, this effect can be suppressed and the device efficiency increased \cite{Reineke2009}. Additionally, the transition from glass to air at a planar interface results in a fanning out of emitted rays, which gets efficiently reduced by the half-sphere. Thus, a strong enhancement of the contrast is possible, as shown in \reffig{fig:Secondary_optics}. In comparison with the bare OLED (see \reffig{fig:Setup_etc}), the forward emission gets more confined and the side emission forms a sharp ring.  The distance between OLED and screen was kept constant during all measurements. Studying the brightness profile, the influence is obtained in a more quantitative way (\reffig[a]{fig:Secondary_optics}). In forward emission, the FWHM decreases to 43$^\circ$ compared to 62$^\circ$ for the reference. In the sideward emission, the contrast between the centre of the spot and the maximum intensity increases from 26\% to 85\%. The corresponding spatial distance between the maxima is $D_{\text{max}}= 82^\circ$ and 67$^\circ$.

Due to the physical shape of an OLED, no emission direction $\varphi$ is preferred over the others leading to the rotational symmetry of the beam as visible in \reffig{fig:Setup_etc} and \reffig{fig:Secondary_optics}. However, some applications may require an asymmetric emission pattern. One way to break this symmetry is to add optical elements without circular symmetry, e.g. a prism, to the beam-shaping OLED. As illustrated in \reffig[a]{fig:Secondary_optics}, the prism covers the pixel completely. Due to its orientation, internal reflections in the y-direction increase, whereas they decrease in the x-direction, compared to the reference. This behaviour leads to a strong change in the brightness distribution on the screen, as is visible in \reffig[c]{fig:Secondary_optics}. Indeed, the beam shape now follows a mixture of the intrinsic radial symmetry of the OLED and the symmetry of the prism. The brightness in the forward emission and the x-direction gives a profile with a FWHM of only 40$^\circ$. For sideward emission, one obtains a contrast value of 71\% and $D_{\text{max}}= 81^\circ$ in the x-direction and a contrast of 60\% at $D_{\text{max}}= 46^\circ$ in the y-direction. 

\subsection*{Conclusion}

We have demonstrated that OLEDs are the first light sources that enable beam-shaping without any further optical elements and real-time continuous adjustment of the spatial brightness distribution. Several challenges were pointed out like the spectral drift of the emission colour especially in sideward emission, due to the curvature of the cavity mode. The use of emitters with spectrally narrower PL spectra would be an adequate response to that point. Furthermore, the emitted colour should be similar in both emission modes. This effect was addressed by utilizing two different emitters of similar spectral distribution but offset PL maxima.

Future work on this topic should focus on the development of green and blue beam-shaping OLEDs to cover the three primary colours. First tests have already been carried out and showed promising results. Nevertheless, the influence of the V($\lambda$) curve for human perception of colours \cite{CIE2006} works contrary to the curvature of the cavity mode, especially in the blue colour regime. On the one hand this requires a far more elaborate sample design, but on the other hand the emission wavelength scales with the layer thickness, which means that process-based deviations in the thickness compared to the simulation have much more impact on the outcoupled spectra. These two facts make it more complex to realize the conceived optical design -- the optical minimum in particular. But having in mind that these are experimental and not theoretical limitations, we are optimistic that further evaluation of this promising concept will be presented in future work, heading towards full colour active beam-shaping light sources.

Clearly, the ultra-thin, area light source character of OLEDs paired with the ability to control their angular emission properties solely through the modulation of the driving signal opens a new opportunity to design and use complex, customized, and adjustable light sources and systems. The interplay between light sources and secondary optics begs further investigation to make use of the full potential of this concept.

\section*{Methods}\vspace{-3mm}
\textbf{Simulation}
All simulations were carried out with in-house developed simulation software. Theoretical details can be found in literature \cite{Furno2012, Neyts1998}. This simulation solves the Maxwell equation for stratified media, assuming the emission layer is infinitesimally thin and thus solving the equation system using Green functions\cite{Sipe1987}. Parameters that have to be set, are the thickness, the refractive index $n(\lambda)$, and the absorption coefficient $k(\lambda)$ of each layer. Furthermore, both the PL-spectrum and the molecular orientation of the emitter has to be known. The software itself has proven functionality with various types of planar devices \cite{Furno2012, Fuchs2015} and even works for corrugated samples \cite{Fuchs2013}.

\textbf{Sample preparation and materials.}
All samples are produced on glass substrates pre-coated with ITO anodes. Organic layers and metals are deposited by thermal evaporation in a UHV chamber at a base pressure of $10^{-6}$-$10^{-7}$~mbar and rates of 0.2-2~\AA/s. The samples are encapsulated to prevent oxygen, water, and dust exposure.

The OLEDs follow the pin-architecture. Two stacked OLEDs are evaporated directly on top of each other, separated by a Au/Ag wetting layer \cite{Schubert2013}. The hole transport layers (HTL) consist of 4 weight percent (wt\%) 2,2'-(perfluoronaphthalene-2,6-diylidene)dimalononitrile (F$_6$-TCNNQ) doped into 2,2',7,7'-Tetrakis-(N,N-dimethylphenylamino)-9,9'-spirobifluoren (Spiro-TTB).
Electron transport layers (ETL) are of caesium doped 4,7-diphenyl-1,10-phenanthroline (BPhen).
Regarding energy barriers\cite{Meerheim2008} and chemical interactions\cite{Scholz2008} the blocking layer materials are chosen from the following pool: BPhen or Aluminum(III)bis(2-methyl-8-quninolinato)-4-phenylphenolate (BAlq$_2$), as hole blocking layer (HBL), 2,2',7,7'-tetrakis-(N,N-diphenylamino)-9,9'-spirobifluorene (Spiro-TAD) or N,N$^\prime$-Di(naphthalen-1-yl)-N,N$^\prime$-diphenyl-benzidine (NPB) for the electron blocking layer (EBL).
Red emission is obtained either by Iridium(III)bis(2-methyldibenzo-[f,h]chinoxalin)(acetylacetonat) (Ir(MDQ)$_2$(acac)) or Tris(1-phenylisoquinoline)iridium(III)  (Ir(piq)$_3$). Both are phosphorescent emitters, which are doped into a matrix of NPB with 10~wt\%, each.
%Green OLEDs provide an emission layer (EML) of Bis(2-phenylpyridine)iridium(III)acetylacetonate (Ir(ppy)$_2$(acac)) or	Tris(2-phenylpyridine)iridium(III) (Ir(ppy)$_3$), doped into a double layer of 4,4$^\prime$,4$ ^{\prime \prime}$-tris-(N-carbazolyl)triphenylamine (TCTA) and 2,2$^\prime$,2$ ^{\prime \prime}$-(1,3,5-phenylen)tris(1-phenyl-1H-benzimidazol) (TPBI). The doping concentration in both cases is 8~wt\%.
%Blue emission layers consist of 2-Methyl-9,10-bis(naphthalen-2-yl)anthracene (MADN), which is doped with 1,5 wt\% 2,5,8,11-Tetra-tert-butylperylene (TBPe).
Details about device architecture and layer thicknesses are provided in the supplementary information, Fig. S1.

\textbf{Characterization.}
Even though the final device is AC driven to enable active beam-shaping, a complete characterization of each subunit is done under DC conditions. Electro-optical performance, as current and luminance against voltage, is measured using a source measure unit (SMU2400, Keithley Instruments) combined with a calibrated spectrometer (CAS140, Instrument Systems GmbH). Angularly-resolved measurements are done using a spectro-goniometer setup that measures the irradiance with a calibrated USB-spectrometer (USB4000, Ocean Optics Inc.). Here, the OLED is mounted on a rotary table and the spectrometer is at a constant distance to the device. The information from this measurement is also used to calculate correct efficiency values, which is in this case indispensable due to the highly non-Lambertian behaviour of the samples.
To get an image of the beam shape another setup has been developed. In this case the OLED illuminates a flat, diffusive screen. Using a digital single-lens reflex camera (EOS-D30, Canon) and subsequent image evaluation tools, the brightness distribution can be analysed. For evaluation, the spectra are flattened using a FFT lowpass filter. Asymmetries in the profiles hint that the screen was not perfectly parallel to the OLED. Therefore, the mean value of both maxima has to be used.

The power supply for the PWM is obtained by an arbitrary waveform generator (33220A, Agilent Technologies) and a subsequent 5~MHz high voltage amplifier (WMA 300, Falco Systems).

\section*{Acknowledgements}
This project has received funding from the European Research Council (ERC) under the European Union’s Horizon 2020 research and innovation programme (grant agreement No 679213). 

\section*{Author contributions}
F.F. carried out the simulation, experiments and evaluation, S.L. supported the device design, S.L. and S.R. helped interpreting the data, M.F. supplied assistance with the simulation, P.Y.A. carried out the experiments with half-sphere and prism. F.F., S.L., and S.R. devised the original concept. All authors jointly wrote the manuscript.

\section*{Additional information}

Supplementary information has been submitted along with the main article.
Correspondence to: Felix Fries, Sebastian Reineke
%\nolinenumbers

\newpage
\clearpage
\thispagestyle{empty}
\mbox{}
\newpage
\setcounter{page}{1}

\vspace*{1cm}

\begin{center}
\Large Supplementary Information to:

\huge
\textbf{Real-time beam-shaping without additional optical elements} 

\vspace{2cm}
\Large
Felix Fries$^\star$,
Markus Fr\"obel,
Pen Yiao Ang,
Simone Lenk,
Sebastian Reineke$^\star$

\vspace{1cm}
\large
Dresden Integrated Center for Applied Physics and Photonic Materials (IAPP),\\ \small Technische Universit\"at Dresden, N\"othnitzer Strasse 61, 01187 Dresden, Germany

\vspace{2cm}
\begin{tabular}{cl}
$^\star$E-Mail: & felix.fries@iapp.de \\
& sebastian.reineke@iapp.de
\end{tabular}
\end{center}\normalsize\clearpage
%\linenumbers*

\section{Device Architecture}
\reffig[a)]{fig:stack_field} shows a detailed scheme of the device architecture. On top of the glass substrate, an ITO electrode is located. Then two pin-OLEDs are stacked on top, separated by a wetting layer electrode of only 10~nm metal. Doped layers are given with the corresponding doping ratio. In the case of BPhen:Cs the doping ratio is chosen to achieve a conductivity of $10 ^{-5}$~S/cm. 
Additionally, the electrical connections are indicated in \reffig[a)]{fig:stack_field}.

\begin{figure}[ht]
	\centering
	\footnotesize
	\includegraphics[scale=0.5]{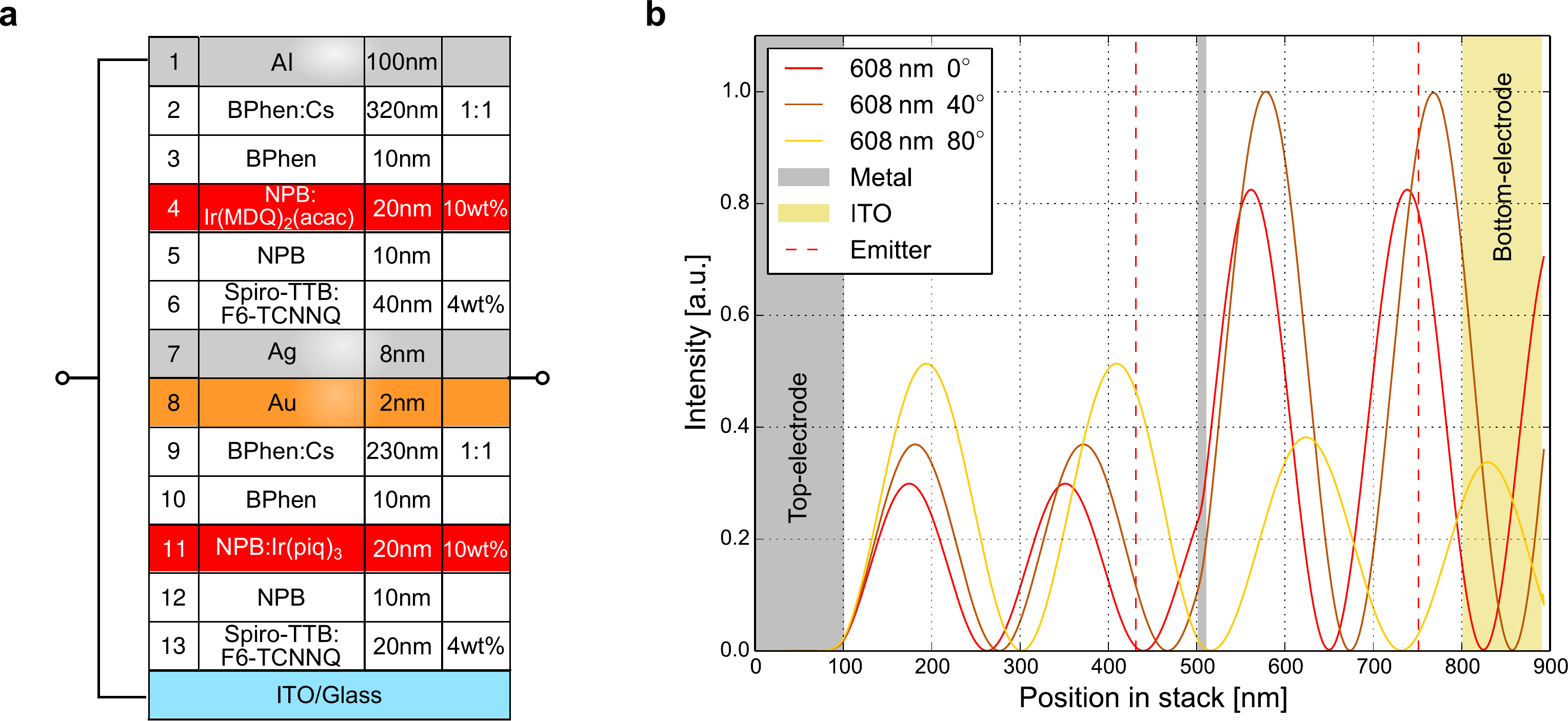}
	\caption{\textbf{Stack Architecture} Detailed scheme of the device architecture. Long names of the materials can be found in the materials section of the main article.}
	\label{fig:stack_field}
\end{figure}

Solving Maxwell's equations for the stratified device, one obtains a distribution of the intensity of the electric field.\cite{SI:Sipe1987,SI:Heavens1991} In the simulation used in this contribution the field is, of course, taken into account. The visualization of the location of the emissive layer within this field helps to interpret the out-coupled spectra. As shown in \reffig[b)]{fig:stack_field}, the emission layer for side emission is located close to a field minimum at 0$^\circ$. By increasing the viewing angle, a local maximum is shifted to the position of the emission layer. Note that for simplicity here the emission zone is assumed to be infinitesimal thin. The emission zone in the forward emission unit faces a local maximum at 0$^\circ$, which increases in intensity with increasing viewing angle up to 35$^\circ$ \ and only then starts to drop.

\clearpage
\section{PWM}

\begin{figure}[ht]
	\centering
	\scriptsize
	\def\svgwidth{1\textwidth}
	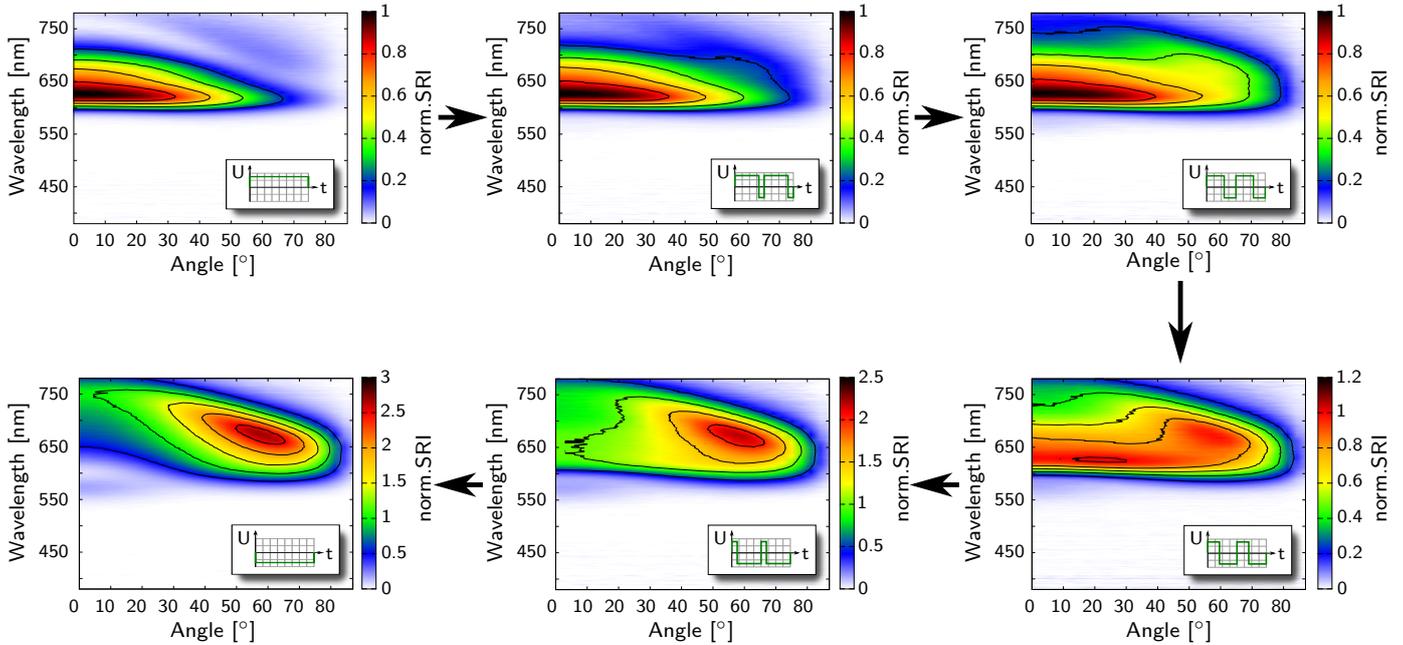
	\caption{\textbf{Spectra under AC conditions} Changing the duty cycle, the spectra continuously change from forward to side emission. Duty cycle increment is 20\%.}
	\label{fig:PWM-spectra}
\end{figure}

\begin{figure}[ht]
	\centering
	\includegraphics[width=0.45\textwidth]{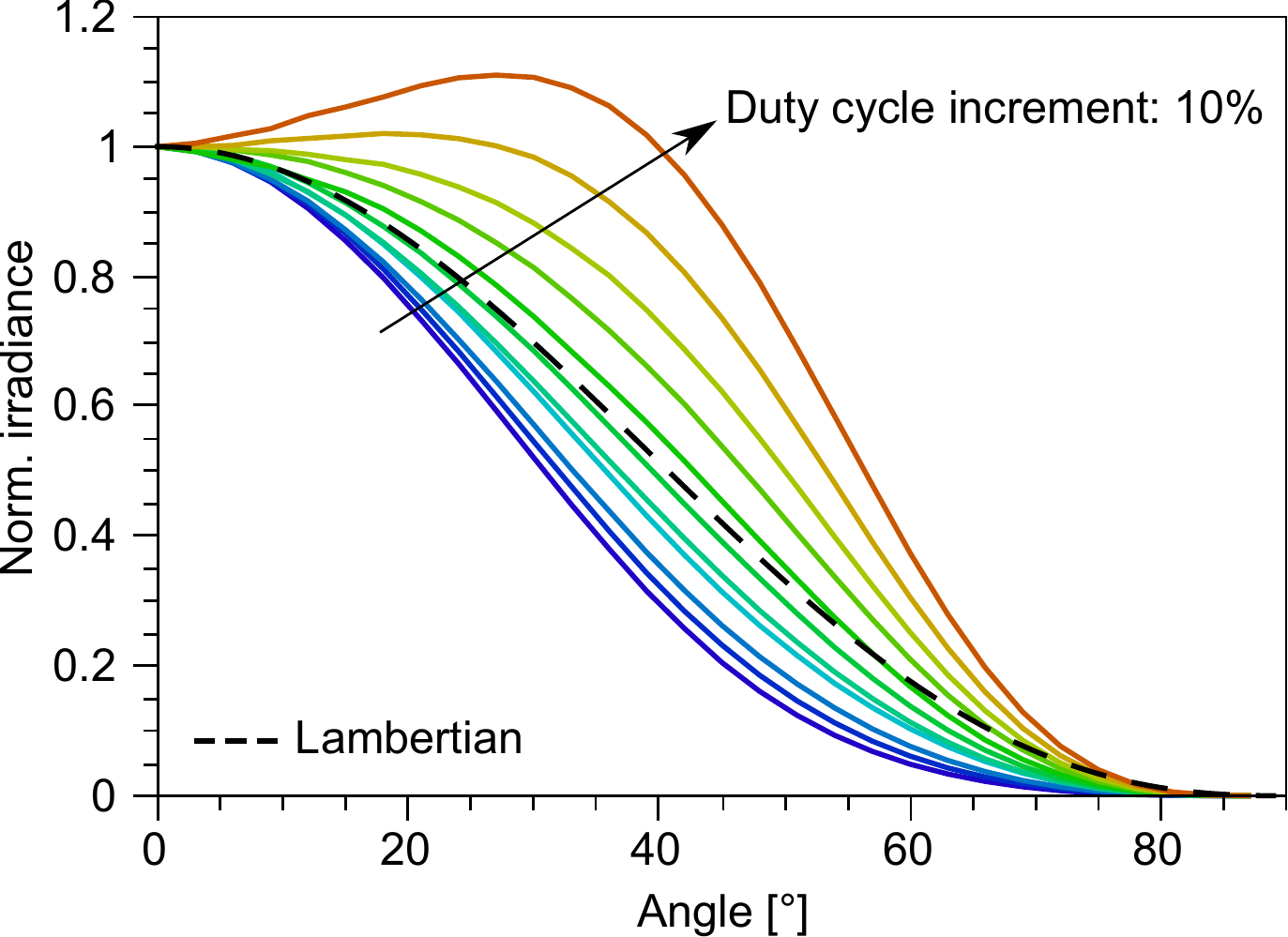}
	\caption{\textbf{Irradiance under AC conditions.} Integrating the spectra for different PWM signals over the wavelength gives the irradiance development from forward to side emission for various duty cycles. For comparison the profile of a Lambertian emitter is shown.}
	\label{fig:PWM-irradiance}
\end{figure}

Using PWM, the ratio of primarily side emission to primarily forward emission can be adjusted. In the main article three examples show how the image of the beam shape can be modified (Fig 1). Of course, this can also be measured with the spectro-goniometer. In \reffig{fig:PWM-spectra}, the angle-dependent spectral radiant intensity (SRI) is presented for different waveforms. From the top-left to the bottom-left the ratio of forward to side emission changes from 100\% duty cycle to 0\% duty cycle in 20\% steps. All the spectra are individually normalized to their maximum value at 0$^\circ$. Integrating each of the spectra in \reffig{fig:PWM-spectra} over the wavelength, one obtains the irradiance development shown in \reffig{fig:PWM-irradiance}. The duty cycle increment here is 10\%.

\clearpage
\section{Comparison of simulated and experimental spectra}

\begin{figure}[ht]
	\centering
	\includegraphics[width=\textwidth]{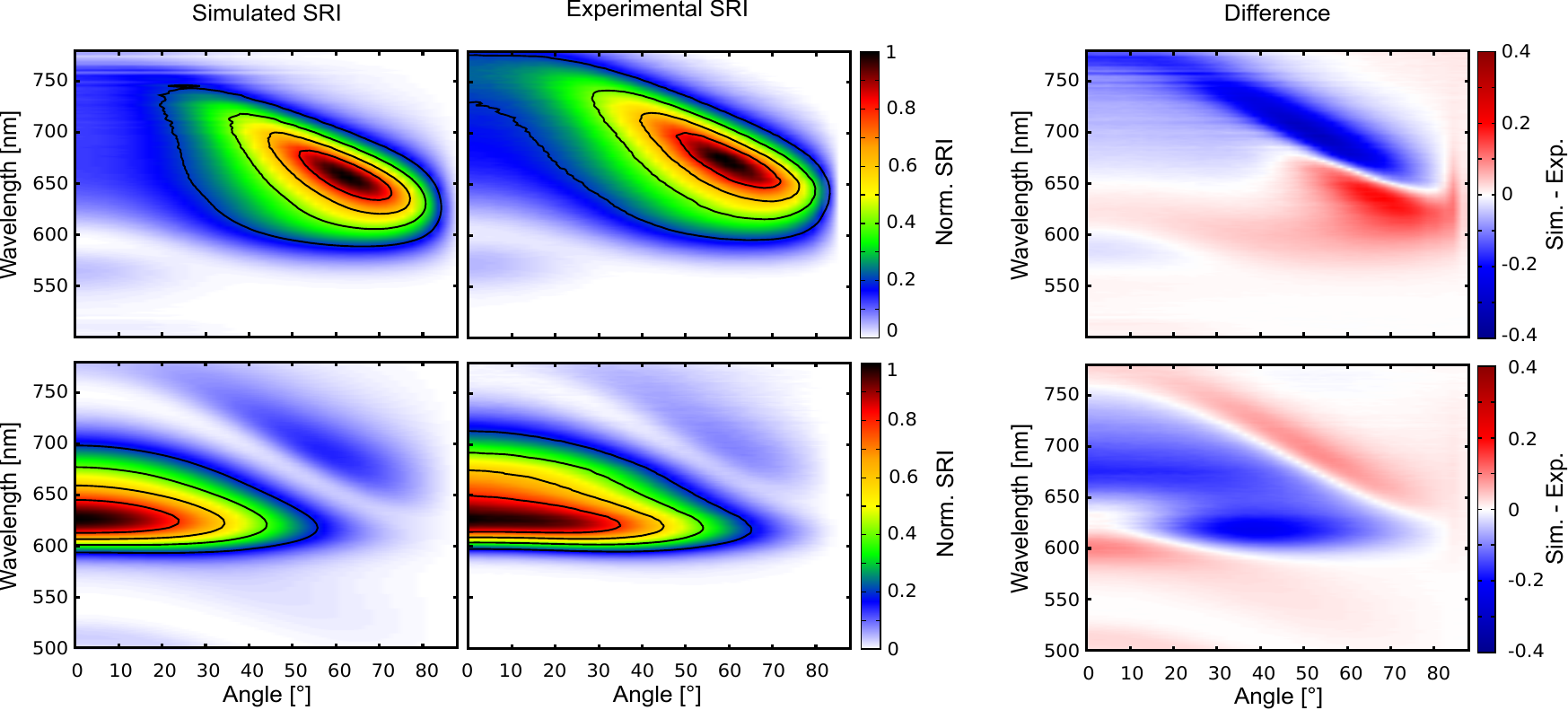}
	\caption{\textbf{Simulated and measured spectra.} The left column shows the simulated spectra of the side-emission unit and the forward-emission unit, respectively. Beside it, in the middle column, the experimentally measured spectra are shown for comparison. The qualitative behaviour is reproduced fairly well. For quantitative analysis, difference spectra were calculated, as shown in the right column.}
	\label{fig:Comparison_exp_sim}
\end{figure}

As shown in \reffig{fig:Comparison_exp_sim}, the simulated spectra predict qualitatively quite well the experimentally measured behaviour. For side emission, low intensity can be seen at 0$^\circ$, rising to a maximum around 60$^\circ$. The forward-emission unit shows strong emission around 608~nm and 0$^\circ$, dropping in intensity fast with increasing viewing angle. Also a parasitic mode around 700~nm and 65$^\circ$ \ is reproduced, even though in the experiment it is much less pronounced.

For a quantitative analysis, the difference between the two spectra (simulation minus experiment) is calculated and shown in the right column. In side emission it becomes clear that the experimental spectra are slightly red shifted compared to the simulation, which hints that the cavity length and thus the resonance wavelength of the resonator is not exactly as expected. In forward emission, differences in the shape are visible that way. The simulation drops faster than the experimental data and again the parasitic mode is more pronounced.

Adjusting the layer thickness of several layers in the simulation may lead to a better agreement and thus could provide further information on the processing procedure.

\clearpage
\section{Gnomonic Transformation}

As written in the main text, the transition between the spectro-goniometer and the planar screen is described by an azimuthal transformation, precisely a gnomonic projection. Treating the OLED as a point source, the emitted light symmetric with respect to the azimuthal angle $\varphi$, the detector at a distance $R$, and its aperture $\diff A$ being much smaller then $R$, which than is equal to the surface element in spherical coordinates:
\begin{equation}\label{SI-eq:dA}
	\diff A = R^2 \sin \theta \diff \varphi \diff \theta \hspace{2mm}.
\end{equation}

Using the same notation as in the main text, the respective area on the flat screen, can be written as the surface element in polar coordinates $\diff A^\prime$, which is at a distance $r$ from the sphere's north pole:
\begin{equation}\label{SI-eq:dAprime}
	\diff A^\prime = r \diff r \diff \varphi \hspace{2mm},
\end{equation}
where the relation between $r$ and $R$ is given as:
\begin{equation} \label{SI-eq:R-Theta}
	r = R\cdot \tan \theta \hspace{2mm}.
\end{equation}

With this the transformation between the two measurement systems is straightforwardly obtained:
\begin{equation*} \diff r = \frac{\partial r}{\partial R} \diff R + \frac{\partial r}{\partial \theta} \diff \theta~ \stackrel{\diff R = 0 \vspace*{1mm}}{=}~ \frac{R}{\cos^2 \theta} \diff \theta \hspace{2mm},
\end{equation*}
\begin{equation}\label{SI-eq:dAprime-zu-dA}
	\Rightarrow\diff A^\prime = \frac{1}{\cos^3 \theta} \diff A \hspace{2mm}.
\end{equation}

With the definition of the irradiance $E$ as the derivative of the photonflux $d\Phi$ with respect to the detecting surface $dA$, one gets:

\begin{equation}
	\Rightarrow E(\lambda)_\text{Goniometer} = \frac{1}{\cos^3 \theta} \cdot E(\lambda)_\text{Screen} \hspace{2mm}.
\end{equation}

As the aperture of the detector is constant, assuming a cylindrical symmetry, rather than a spherical, appears to be appropriate. This results in:\\
\begin{equation}\label{SI-eq:Irradiance-Trafo}
	E(\lambda)_\text{Goniometer} = \frac{1}{\cos^2 \theta} \cdot E(\lambda)_\text{Screen} \hspace{2mm}.
\end{equation}

\begin{figure}[hb]
	\centering
	\includegraphics[width=\textwidth]{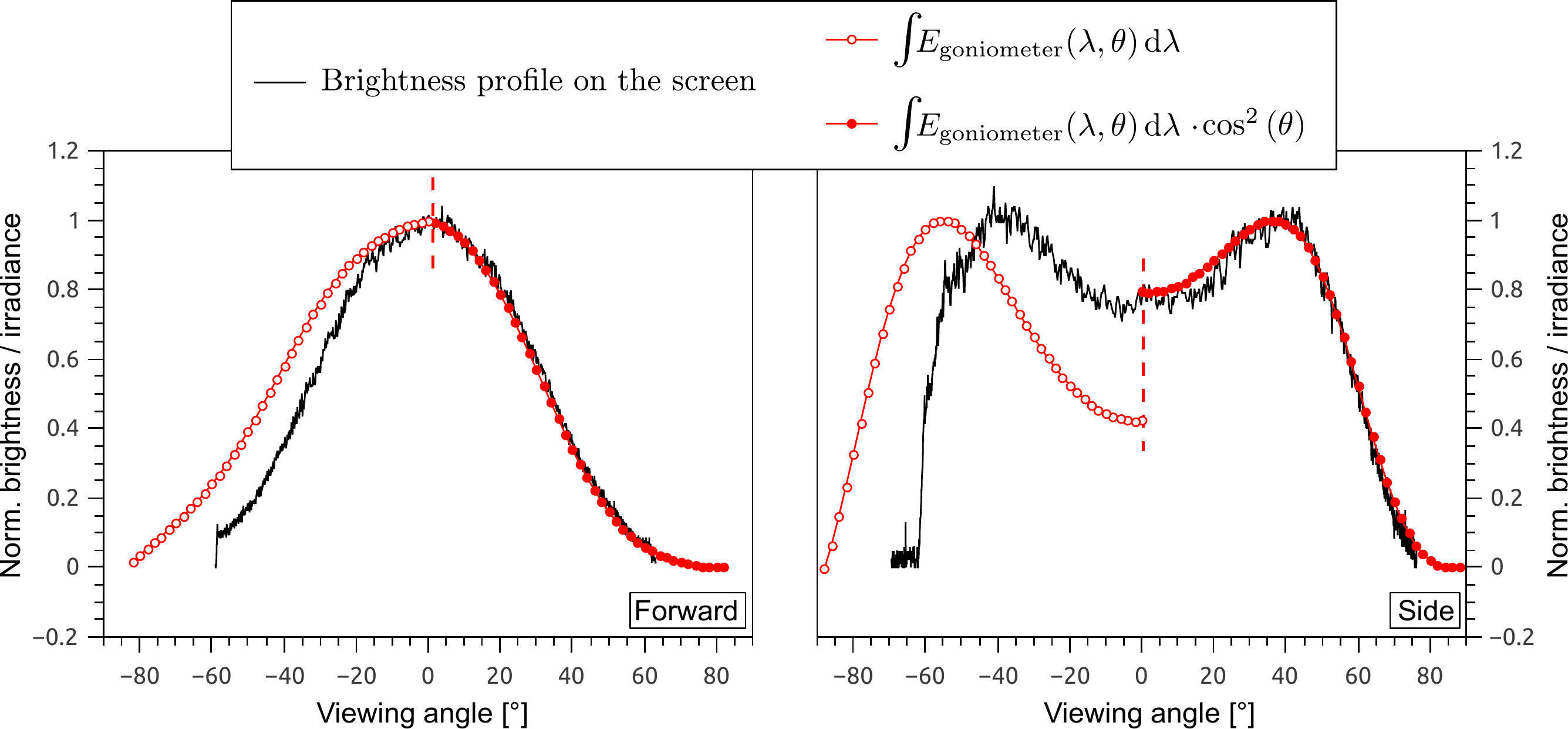}
	\caption{\textbf{Brightness profiles under different conditions.} The black lines show the brightness distribution on the screen. The abrupt drop around -60$^\circ$ \ in side emission is caused by the shadow of the sample holder. Red lines with white dots show the irradiance measured at the spectro-goniometer. Applying a transformation factor $\cos^2(\theta)$ on the latter, both measurements coincide as shown in red lines with solid dots.}
	\label{fig:Trafo_profile}
\end{figure}

In \reffig{fig:Trafo_profile}, the measured brightness distribution on the screen (black) is compared to the integrated irradiance, obtained using the spectro-goniometer (non-filled red dots). Especially in the side emission, a strong difference in the shape is obvious. The contrast between 0$^\circ$-emission and the maximum is way higher in the case of the goniometer. Furthermore, the profile peaks around 60$^\circ$ \ compared to 40$^\circ$ \ on the screen. When multiplying the goniometer profile with $\cos^2(\theta)$, those data resemble the black line very well. Both the contrast and the maximum angle is reproduced. In forward emission direction, the drop of intensity coincides, as well. Actually, it is surprising that the cosine of the angle to the power of two gives the best results, as a power between two and three was expected to model the transformation most realistically. Obviously, some of the assumptions made are not completely fulfilled. For example, in Fig.~4b) it can be seen, that the sample may not be treated as a point source. Nevertheless, accepting this factor two, the profiles of the measurements are highly comparable.

\begin{figure}[hb]
	\centering
	\includegraphics[width=\textwidth]{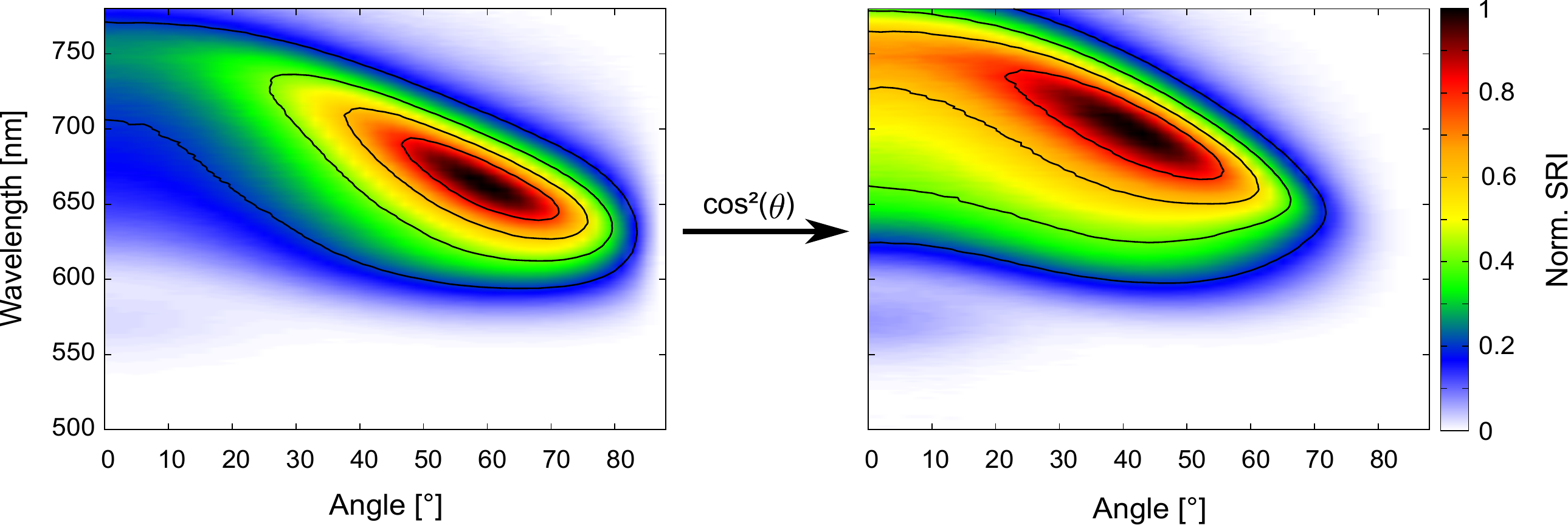}
	\caption{\textbf{Spectral transformation.} Applying the azimuthal transformation on the spectral radiant intensity, the contrast between 0$^\circ$ \ and maximum emission gets reduced.}
	\label{fig:Trafo_spectra}
\end{figure}

With this information the expected beam shape on a screen can be calculated from the spectro-goniometer data or even from simulation data, which opens further possibilities for sample optimization. The first step to be done is to transform the spectra. In \reffig{fig:Trafo_spectra}, this is shown for the experimental data. Of course, this transformation leads to an enhancement of low-angle parts, shifting the maximum to a lower angle. Subsequently, the CIE-XYZ coordinates are calculated for each angle, using the CIE2006 colour-matching functions\cite{SI:CIE2006}.

\vspace*{1cm}

\clearpage
The last step towards a true colour image is a linear transformation ($\mathbf{\mathcal M}_{3\times3}$) from the CIE-XYZ to a displayable RGB colour space. In case of the sRGB space with D65 as reference white point this reads:\cite{SI:Hunt2004}

\begin{equation}
	\left( \begin{array}{c}
		R \\
		G \\
		B
	\end{array} \right)
	=
	\left( \begin{array}{ccc}
		3.24 & -1.54 & -0.50 \\
 		-0.97 & 1.88 & 0.04 \\
		0.06 & -0.20 & 1.06 \\
	\end{array} \right)
	\left( \begin{array}{c}
		X \\
		Y \\
		Z	
	\end{array} \right)
\end{equation}

This gives an 1D-array containing one RGB-tuple for each angle, which has to be multiplied with the calculated integrated irradiance distribution (equation~\refeq{SI-eq:Irradiance-Trafo}). Note that with this, the ratio between the R,G, and B coordinates is unchanged which results in a sole change of the brightness, but not of the respective colour. Assuming again azimuthal symmetry this 1D-colour-brightness distribution can be expanded to a 2D-colour-image.

\FloatBarrier

\end{document}